\documentclass[final,3p,times,twocolumn]{elsarticle}

\usepackage{amssymb}
\usepackage{graphicx}
\usepackage{epstopdf}
\usepackage{wrapfig}
\usepackage{xcolor}
\usepackage{footnote}
\begin{document}

\begin{frontmatter}

\title{ Fusion  of $^{12}$C+$^{28}$Si at deep sub-barrier energies\\}
\author[1]{\small A.M. Stefanini}
\affiliation[1]{organization={INFN, Laboratori Nazionali di Legnaro},
            city={Legnaro},
            postcode={I-35020},
            country={Italy}}
            
\author[2,3]{G. Montagnoli}
\affiliation[2]{organization={Department of Physics and Astronomy, University of Padua},
            city={Padua},
            postcode={I-35020},
            country={Italy}}
  \affiliation[3]{organization={INFN, Section of Padua},
            city={Padua},
            postcode={I-35020},
            country={Italy}}  
                    
\author[2,3]{M. Del Fabbro}
 
\author[1]{A. Goasduff}
  
\author[2,3]{P. A. Aguilera Jorquera}
 
\author[2,3]{G. Andreetta}

\author[1,2]{F. Angelini}
    
\author[2]{L.V. D'Auria}
       
\author[1]{M. Balogh}

\author[3]{D. Bazzacco} 

\author[2,3]{ J. Benito}
 
\author[4]{G. Benzoni}
\affiliation[4]{organization={INFN, Section of Milan},
            city={Milan},
            postcode={I-20133},
            country={Italy}}
            
% ...

\begin{abstract}
{\small The existence of fusion hindrance is not well established in light heavy-ion systems. 
Studying slightly heavier cases allows extrapolating the trend to light systems of astrophysical interest. Fusion of $^{12}$C + $^{28}$Si has been measured down to deep sub-barrier energies, using $^{28}$Si beams from the XTU Tandem accelerator of LNL on thin $^{12}$C targets. The fusion-evaporation residues were detected by a detector telescope following an electrostatic beam separator, and coincidences between the $\gamma$-ray array AGATA and segmented silicon detectors DSSD were performed, where the evaporated light charged particles were identified by pulse shape analysis.
Fusion cross sections have been obtained in the wide range $\sigma\approx$150 mb -- 42 nb. Coupled-channel (CC) calculations using a Woods-Saxon potential reproduce the data above $\simeq$0.1 mb. Below that, hindrance shows up and the CC results overestimate the cross sections which get close to the one-dimensional potential tunnelling limit. This suggests that the coupling strengths gradually vanish, as predicted by the adiabatic model. The hindrance threshold follows a recently updated phenomenological systematics.}
\end{abstract}
{}

\end{frontmatter}

\section{Introduction}


The phenomenon of low-energy hindrance in heavy-ion fusion is a topic of ongoing experimental and theoretical interest. It was first observed for the system $^{60}$Ni + $^{89}$Y~\cite{Jiang:2002aa}, and it is experimentally recognised by the increasing logarithmic slope of the excitation function (or by a maximum of the S-factor) showing up at low energies. 

From the theoretical point of view, extending the standard coupled-channels (CC) model to describe the hindrance effect is a theoretical challenge~\cite{Misicu:2007aa,Esbensen:2007aa,Ichikawa:2007aa,Ichikawa:2015aa}. A few years ago, Simenel et al.~\cite{Simenel:2017aa,Simenel:2018aa} pointed out that the Pauli exclusion principle hinders the overlap of the two colliding nuclei, thereby influencing the ion-ion potential. As a consequence, the Coulomb barrier
turns out to be thicker and higher, and low-energy fusion hindrance is produced. 

For medium-mass systems, where the fusion $Q$-value is negative, hindrance has been systematically observed at various cross-section levels and with different features.
In the case of light systems, the S-factor maximum becomes broader and the hindrance threshold is more difficult to recognise. Such light systems have positive fusion $Q$-value, implying that the existence of an S-factor maximum is not algebraically necessary~\cite{Jiang:2006ab}. Indeed, the fusion hindrance is neither well-established nor understood in those cases. This creates uncertainties when extrapolating their trend to astrophysical energies, where it may influence the reaction rates in stellar environments~\cite{Gasques:2007aa}.

In more detail, $^{12}$C + $^{16}$O~\cite{Fang:2017aa} and $^{12}$C +  $^{12}$C (see~\cite{Tan:2020aa,Fruet:2020aa} and Refs. therein), the existence and the features of that phenomenon are obscured by the presence of several low-energy oscillations of the S-factor. 
The case of $^{12}$C +  $^{13}$C~\cite{Zhang:2020aa} is completely different because no oscillations and no hindrance have been observed. As the energy decreases, the S-factor tends to develop a maximum; however, it then increases again.

In this respect, we point out that a very recent theoretical study~\cite{Uzawa:2025aa} indicates the absence of the hindrance effect in both $^{12}$C +  $^{12,13}$C, where a reexamination of their low energy behaviour is presented. Uzawa and Hagino propose a modified fitting procedure in that energy range, and show that the resulting astrophysical S-factors do not show any hindrance within the range of error bars for both systems.


The oscillating behaviour of $^{12}$C + $^{16}$O was associated to the elastic $\alpha$-transfer channel. Alternatively,  it 

\footnote{{\textdagger} deceased} 

was recently suggested to arise from quasi-molecular resonances~\cite{Fang:2017aa}.
In $^{12}$C +  $^{12}$C, the pronounced oscillations were initially attributed to quasi-molecular resonances~\cite{Bromley:1960aa,Erb:1980aa} giving rise to a widespread debate about their origin linked to the $\alpha$-like nature of the two nuclei (see ~\cite{Freer:2018aa,Diaz:2018aa} and Refs. therein). More recently, it has been proposed that the oscillations are of resonant origin, caused by the low-level density of the compound nucleus $^{24}$Mg in the relevant excitation energy range~\cite{Jiang:2013aa}.

 \begin{figure}[ht]
\includegraphics[width=7cm]{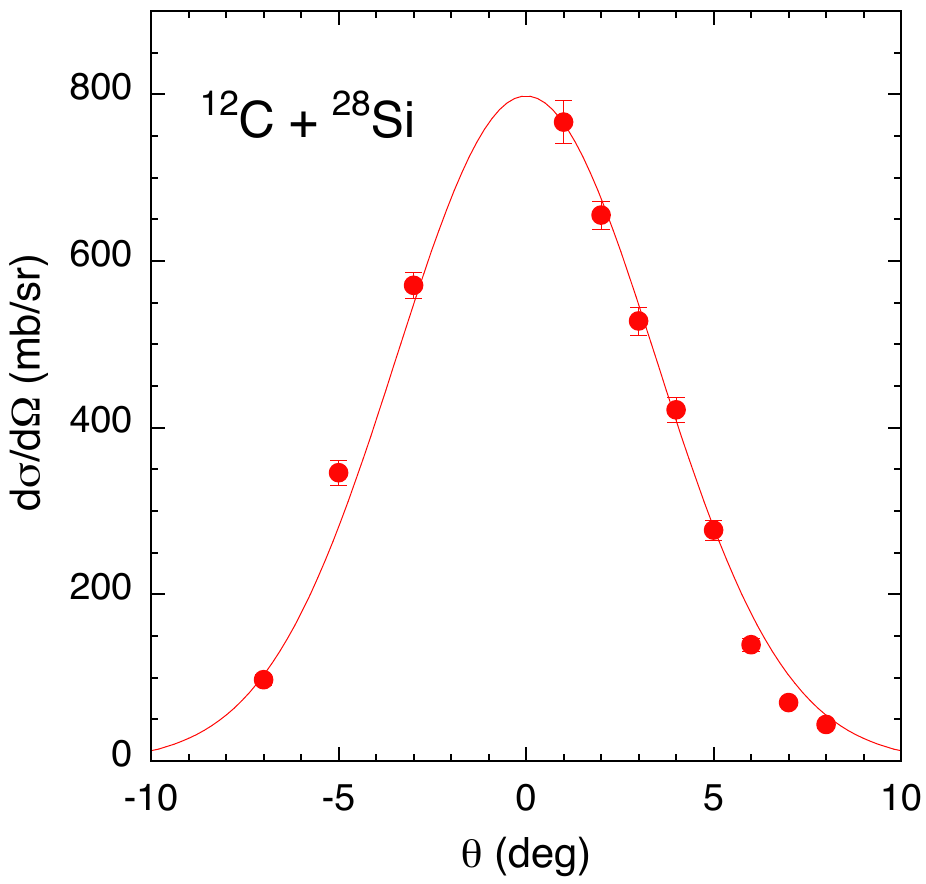}
\caption{\it Angular distribution of ER measured at E$_{lab}$= 43 MeV, together with the Gaussian fit, see text. }
\label{lazero}
\end{figure}

Because of these various features, it was proposed to study slightly heavier systems~\cite{Jiang:2014aa}, to provide a reliable starting point for the extrapolation to lighter ones in the expected energy range of hindrance.
We recently investigated the cases $^{12}$C + $^{24,26}$Mg~\cite{Montagnoli:2022aa,Stefanini:2023aa} and $^{12}$C + $^{30}$Si~\cite{Montagnoli:2018aa}. 
It was observed that the hindrance energy thresholds for these systems follow the empirical estimate of Ref.~\cite{Jiang:2009aa}, updated from the original formulation on the basis of those recent results and the newly published ones on $^{16}$O + $^{48}$Ca~\cite{Stefanini:2025aa}.
The threshold for $^{12}$C + $^{24}$Mg corresponds to the rather large fusion cross section $\sim$0.9 mb, however, the observed oscillations and the corresponding experimental uncertainties make that identification somewhat doubtful for this system.

In this work, we present the results of the experimental study of $^{12}$C + $^{28}$Si, intending to identify its hindrance threshold. 
This information will be essential to recognise how adequate is the extrapolation of the trend to the  astrophysically relevant cases, the more so if cross-sections far below the barrier will be available. Previous fusion data on $^{12}$C + $^{28}$Si are available only above the barrier~\cite{Jordan:1979aa,Gary:1982aa,Lesko:1982aa,Nagashima:1982aa,Harmon:1986aa}.

We have used two different setups to investigate the fusion excitation function far below the barrier, where the cross sections for  $^{12}$C + $^{24}$Mg, $^{30}$Si seem to reach the one-dimensional potential tunnelling limit.  The excitation function of $^{12}$C + $^{28}$Si was measured at Laboratori  Nazionali di Legnaro (LNL) INFN, using the electrostatic deflector setup~\cite{Stefanini:2010aa} for the high-energy part, and the AGATA $\gamma$-ray spectrometer~\cite{Valiente:2023aa} and segmented silicon detectors (DSSD) for the identification of evaporated light charged particles, at the lower energies.

 Sect.~II describes the experimental set-ups and methods of data analysis, and shows 
the results that are then compared in Sect.~III with  CC calculations. Systematic trends are discussed in Section~IV comparing with near-by systems and concerning the astrophysical aspects of the results as well. Section~V presents the conclusions of the present work.

\section{Experimental set-ups}
\label{II}

The $^{28}$Si beams of the XTU Tandem at LNL were employed, with currents 15-30 pnA, in the energy range 29.5-54 MeV.
Thin $^{12}$C targets $\sim$50 $\mu g/cm^2$ were used, with isotopic 
enrichment of 99.8$\%$, to minimise the beam energy corrections 
and straggling effects that may increase the unwanted background. 
In the measurements with the electrostatic separator set-up of LNL, the evaporation residues (ER)
were detected using a  $\Delta E-E$ gas-silicon detector and large position-sensitive micro-channel plates (MCP) detectors. 
The beam control and yield normalisation to the Rutherford cross section were ensured by four silicon detectors placed at $\theta_{lab}$=16$^o$ in the scattering chamber (see Ref.~\cite{Stefanini:2010aa} for further details).

The ER angular distribution was measured at E$_{lab}$= 43 MeV in the $\theta_{lab}$ range from -7$^o$ to +8$^o$, and it is reported in Fig.~\ref{lazero}. It is well fitted by a single Gaussian curve (red line),  and it allowed us to extrapolate its shape to the other energies where the fusion yield was measured at only $\theta_{lab}$=2$^o$ (or 3$^o$ at low energies). The fusion cross section was obtained by integrating that distribution. 
Standard PACE4 calculations~\cite{Gavron:1980aa} anticipate that the shape of the angular
distribution does not appreciably vary with energy in the measured range. This has been validated by several previous measurements (see e.g. Refs.~\cite{Montagnoli:2013aa,Montagnoli:2018aa}).
The systematic error on the cross-section scale is estimated to be
$\pm$7-8$\%$ as in previous experiments with that setup~\cite{Stefanini:2010aa}. 

\begin{figure}[t]
\centering
\includegraphics[width=7.5cm]{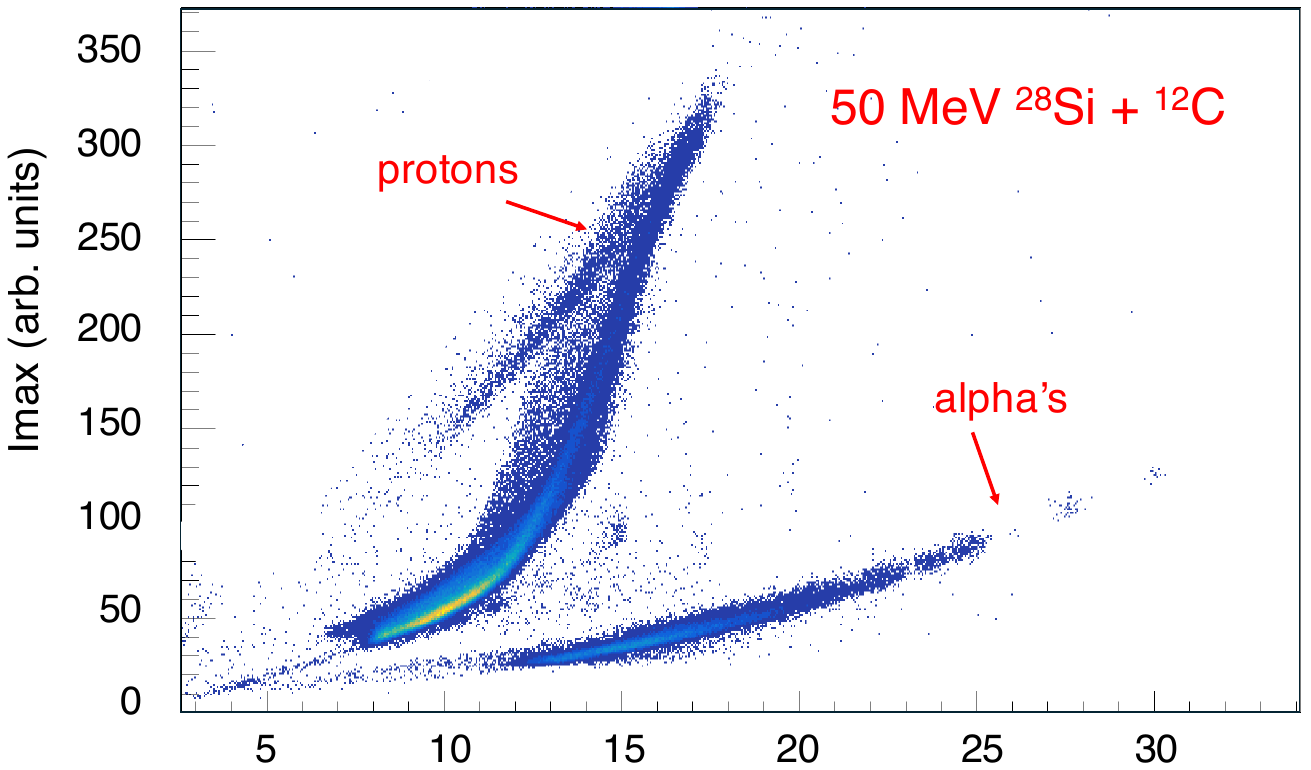}
\includegraphics[width=7.5cm]{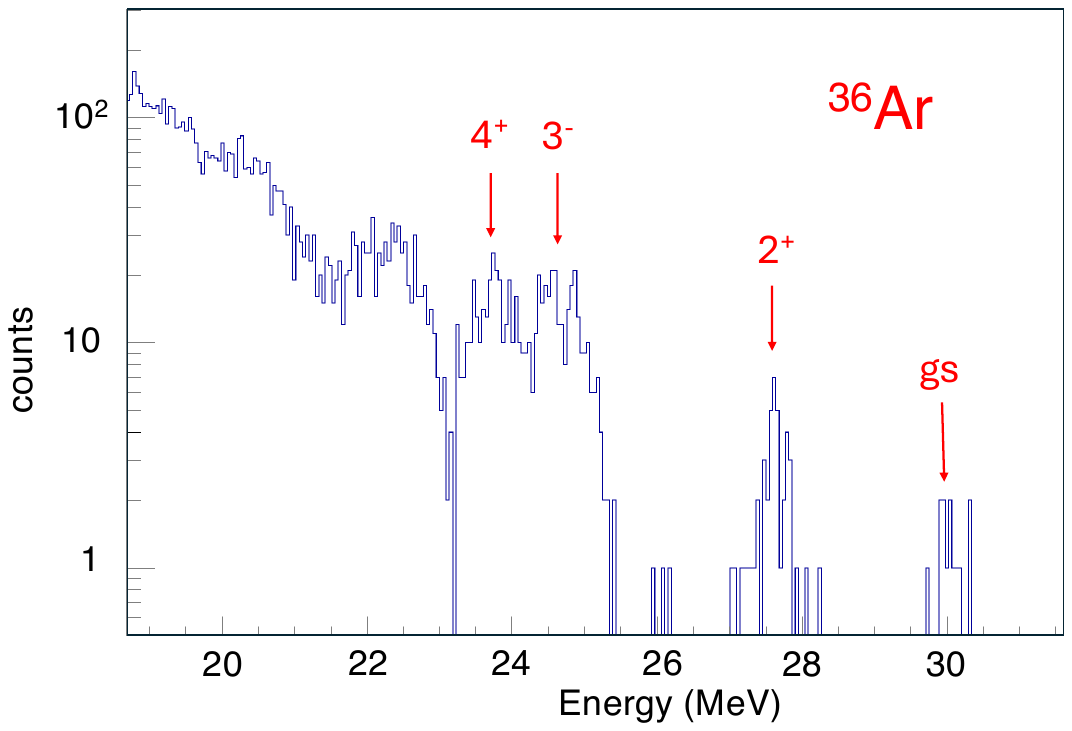}
\caption{\it (top panel) Events detected in an intermediate ring ($\theta\simeq$29$^\circ$)  of the forward DSSD. Protons and $\alpha$ particles are identified through pulse shape analysis ($psa$), plotting the maximum of the signal derivative (I$_{max}$) vs their energy E. (bottom panel) Energy spectrum of the corresponding $\alpha$ particles. Several particle groups populating different states in the $^{36}$Ar residual nucleus can be observed.}
\label{psa} 
\end{figure}

Following the technique introduced by Jiang et al.~\cite{Jiang:2012aa}, we extended the fusion excitation function down to very small cross sections, using
the $\gamma$-ray tracking spectrometer AGATA~\cite{Valiente:2023aa} and two annular Double Sided Silicon strip Detectors (DSSD) (4" diameter) placed 5 cm upstream and downstream of the target.  Their thicknesses were 0.5 and 1.5 mm, respectively, and have been used to detect coincident events between the evaporated light charged particles and the prompt $\gamma$-rays emitted from the various residual nuclei (see Fig. 2 of Ref.~\cite{Brugnara:2025aa}).

The measurements were performed at four $^{28}$Si beam energies with this setup, i.e., E$_{lab}$ = 50, 34 MeV, to overlap with points taken with the electrostatic deflector, and at the very low energies of 31 and 29.5 MeV.

\begin{figure}[t] 
\includegraphics[width=7.5cm]{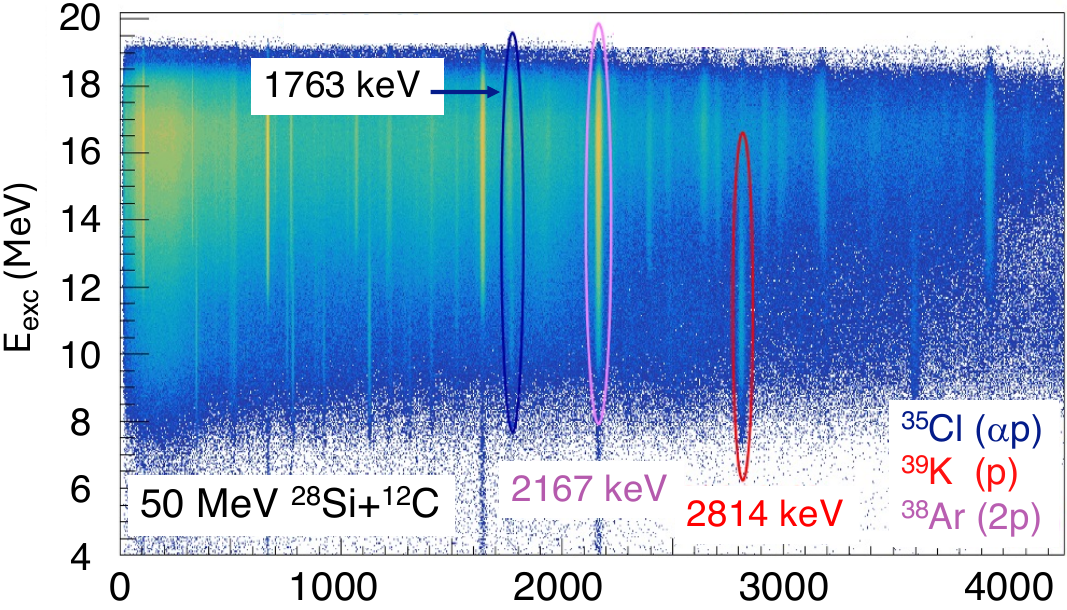}
\includegraphics[width=7.5cm]{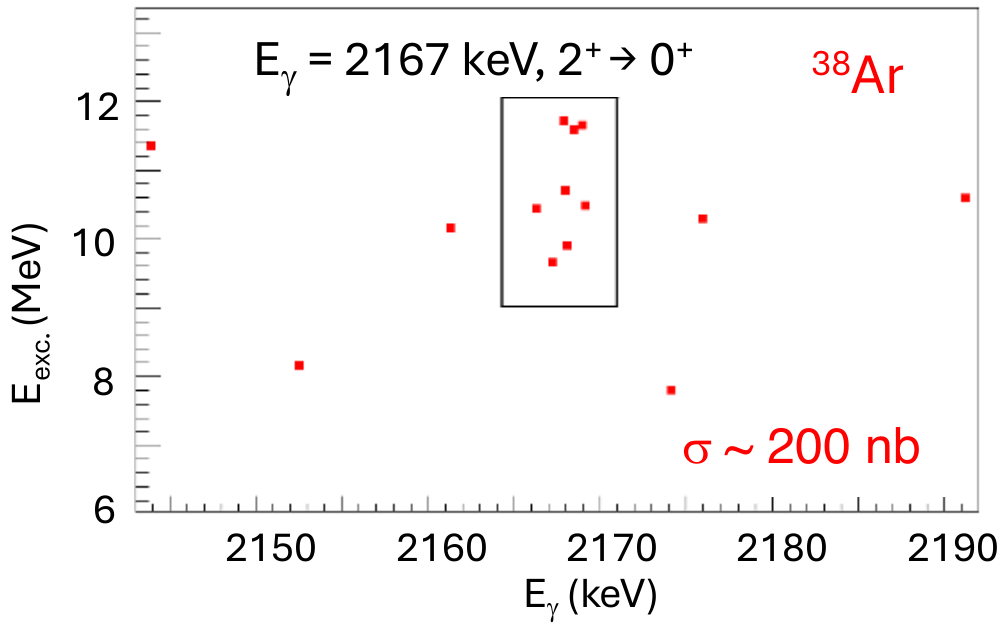}
\caption{\it (top panel) Two-dimensional excitation energy  vs Doppler-corrected $\gamma$-ray energy spectrum for proton events at E$_{lab}$=50 MeV. (bottom panel) Zoomed matrix for the 2p evaporation channel $^{38}$Ar excitation energy ($E_{exc.}$) vs $\gamma$-ray energy at E$_{lab}$=31 MeV, where the few fusion events are clearly identified. The total fusion cross section at this energy is (296$\pm$113) nb.}
\label{psa2}
\end{figure}

Nickel absorbers of calibrated thickness  were placed in front of the two DSSD (15$\mu$m for the forward one, and 2$\mu$m for the backward one)  to stop the scattered beam, target recoils, and the electrons coming from the target. Particle identification by pulse shape analysis was made possible by installing both DSSD with the ohmic side facing the target~\cite{Leneindre:2013aa}. Fig.~\ref{psa} (top panel) shows the good separation obtained between evaporated protons and $\alpha$-particles down to rather low energies.
The direct population of states in the exit channels can then be observed, fixing an emission angle, by projecting this matrix onto the energy axis. This is shown in the bottom panel of Fig.~\ref{psa}, for the $\alpha$-particles selected in the top matrix.

The energy of the evaporated particle, when associated with its emission angle, yields the total excitation of the system. By correlating this excitation energy with the $\gamma$-ray energy,  one can identify the events belonging to a certain evaporation channel. Fig.~\ref{psa2} shows this correlation matrix obtained at E$_{lab}$ = 50 MeV, where all events detected by the two DSSD have been considered (top panel).
This representation is useful even at very low energies, as shown in the bottom panel of the same figure for E$_{lab}$=31 MeV, where the \underbar{eight} fusion events populating $^{38}$Ar by 2p evaporation (corresponding to the indicated cross section of $\simeq$200 nb) are very cleanly identified. 
In the measured energy range, the evaporation channels observed in coincidence events were $1p$, $2p$, $1p1n$, $1\alpha$ and $1\alpha 1p$. 

For each experimental run/energy, the ER level schemes  
provided us with the number of  $\gamma$-particle coincidence events associated to each $\gamma$ transition feeding the corresponding ground states. Subsequently, those yields were normalised using 1) the AGATA efficiency vs $\gamma$-ray energy in the used geometry, obtained by a measurement with a $^{152}$Eu performed just after the experiment, and 2) the DSSD detectors' efficiencies. These are determined by their angular coverage ($\approx$26$\%$ of 4$\pi$), by kinematics and by the electronic thresholds. The sum of all such normalised yields of coincident events, plus those directly feeding the ER ground states (see e.g. Fig.~\ref{psa}),  is then proportional to the fusion cross section measured in the considered run. 

The normalisation between different runs was ensured by two 50 mm$^2$ silicon monitor detectors, installed at $\theta_{lab}$=12$^\circ$ at around 50 cm from the target. Finally, the absolute cross section scale was fixed, taking as reference the energy points at E$_{lab}$=50,34 MeV that was measured also using the electrostatic deflector set-up, whose absolute efficiency is well known for the present system (and several others). Normalising this way the excitation function to the electrostatic deflector results, takes into account also the possible contribution of pure neutron evaporation channels that are not obviously observed in the AGATA-DSSD coincidences, and actually not even in the $\gamma$-ray singles spectra at any energy.

\begin{figure}[t]
\vspace{5 mm}
\includegraphics[width=7.7cm]{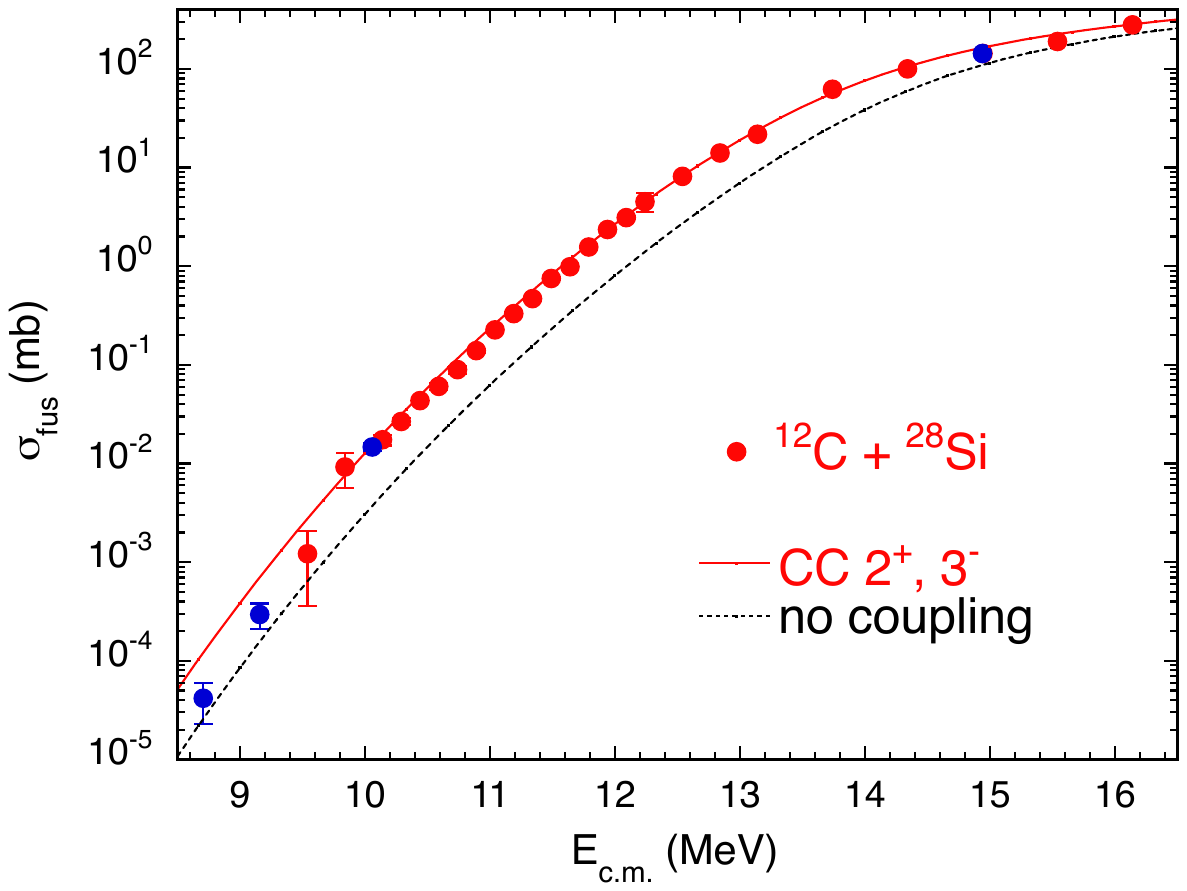}
\caption{\it Measured excitation function of $^{12}$C + $^{28}$Si, compared to CC calculations. The red dots are the points measured with the electrostatic deflector set-up. The blue dots refer to the measurements performed the AGATA + DSSD combination. See text for more details.}
\label{laprima}
\vspace{-0.45cm}
\end{figure}


\section{Coupled-channels calculations}
\label{III}
The fusion excitation function was calculated with the CCFULL code~\cite{Hagino:1999aa} using the coupled-channels (CC) formalism employed in several heavy-ion fusion reactions analyses in recent years. CCFULL takes into account channel couplings to all orders and uses the so-called rotating frame~\cite{Tanimura:1987aa} or isocentrifugal~\cite{Gomez:1986aa} approximation, which considerably reduces the number of channels, thus simplifying the calculations.
A  Woods-Saxon potential with parameters V$_o$=44.6 MeV, r$_o$=1.06 fm and $a$=0.61 fm was used to fit the experimental data near the barrier.

In the calculation, $^{12}$C was considered inert, while 
the lowest quadrupole and octupole excitations of $^{28}$Si were included, with the adopted deformation parameters $\beta_2$=--0.41~\cite{Raman:2001aa} and 
$\beta_3$=0.40~\cite{Kibedi:2002aa}, respectively. The quadrupole deformation parameter of $^{28}$Si is negative, since this nucleus is oblate~\cite{Stone:2005aa}. The mutual excitation of both 2$^+$ and 3$^-$ states was neglected because of its very high excitation energy.

We show in Fig.~\ref{laprima} the obtained excitation function, where the quoted errors are only statistical uncertainties, together with the results of CC calculations. The high-energy points are in agreement with previous results obtained by Jordan et al.~\cite{Jordan:1979aa}.
 The experimental cross sections are well reproduced by the calculation down to $\approx$10 MeV. Below this energy, we have clear evidence of the hindrance phenomenon, and that the three lowest energy points approach and appear to follow the no-coupling limit.

To be noted that the low-energy behaviour of $^{12}$C + $^{28}$Si has been evidenced, even if the experimental uncertainties are rather large, only thanks to the measurements performed with the AGATA spectrometer. 
 
\begin{figure}[ht]
\includegraphics[width=8cm]{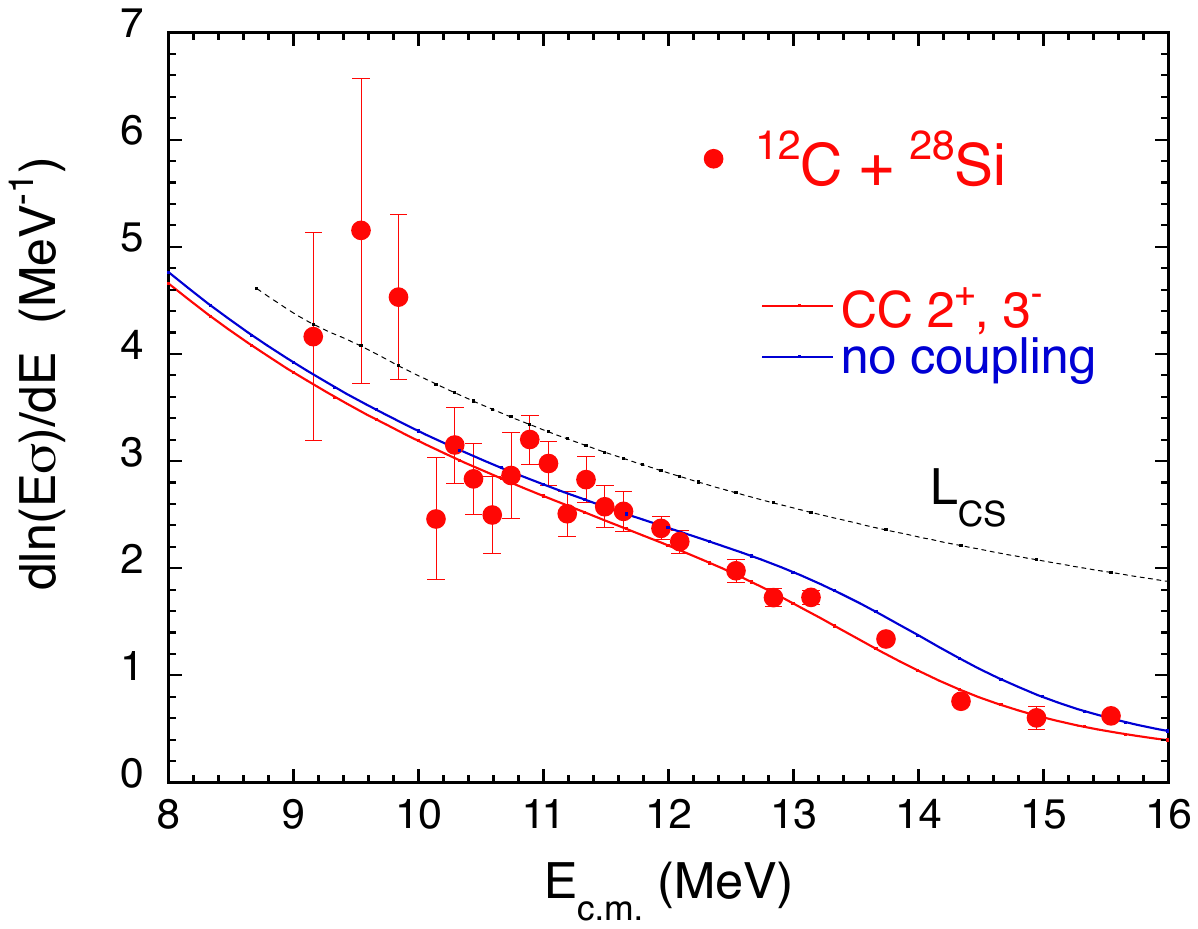}
\caption{\it  The logarithmic derivative of the energy-weighted excitation function for $^{12}$C + $^{28}$Si, compared with the results of CC and no-coupling calculations. The L$_{CS}$ line represents the slope expected for a constant S-factor, for each energy.}
\label{slope0}
\end{figure}

Fig.~\ref{slope0} shows the logarithmic derivative (slope) of the excitation function obtained from the measured cross sections, as the incremental ratio of two near-by points. 
The slope increases with decreasing energy and touches the L$_{CS}$ line at around 11 MeV. A decrease is then observed, followed by a pronounced crossing at $\simeq$10.1 MeV. This is the energy that we can associate to the hindrance threshold in $^{12}$C + $^{28}$Si.

In the same figure, we report the results of the CC calculations and the no-coupling limit. The theoretical curves yield a flat trend for the slope in the plotted energy range, as actually expected. 
They are close to each other, as a consequence of the rather high energies of the $^{28}$Si coupled excitations. The experimental trend is well reproduced by the calculations; however, not where hindrance shows up (and possibly around 11 MeV). This is expected, since one knows that a potential of WS shape is not able, in general, to fit cross sections in the hindrance region~\cite{Misicu:2007aa}.

\section{Comparison with nearby systems}
\label{IV}

\begin{figure}[ht]
\includegraphics[width=8cm]{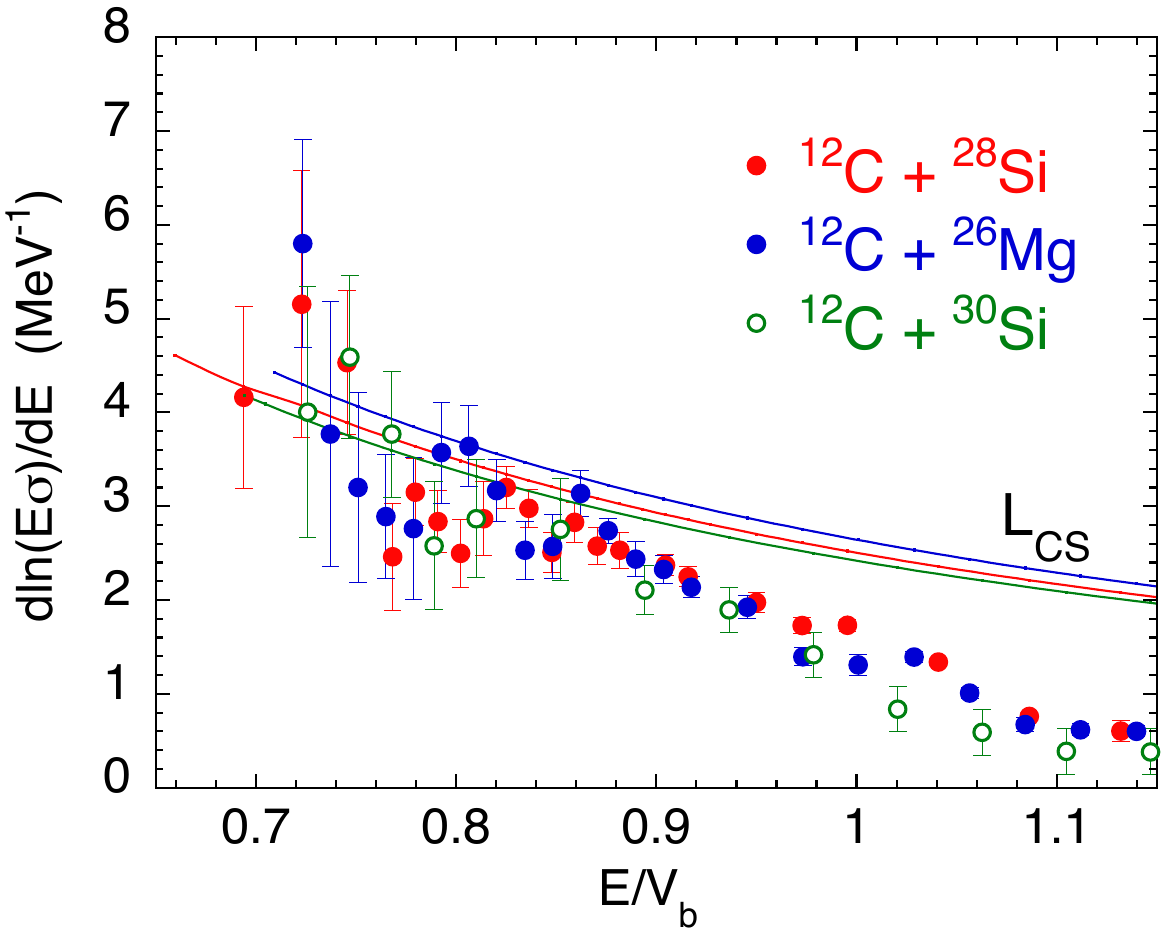}
\caption{\it  Logarithmic derivatives of the energy-weighted excitation functions for $^{12}$C + $^{28}$Si (in red),  $^{12}$C + $^{26}$Mg (blue dots)~\cite{Stefanini:2023aa} and $^{12}$C + $^{30}$Si (open green dots). The three L$_{CS}$ lines are close to each other.}
\label{slope}
\end{figure}

We refer to Fig.~\ref{slope}, where the abscissa is the energy with respect to the Coulomb barrier produced by the Aky\"uz-Winther (AW) potential~\cite{Akyuz:1981aa}. We note the similarity between the logarithmic derivatives of the three systems shown there. With decreasing energy, the three cases exhibit small oscillations followed by a larger slope increase that we associate with the onset of hindrance, at similar $E/V_b$ values.  $^{12}$C + $^{24}$Mg~\cite{Montagnoli:2022aa} (not shown here), has an analogous trend, but the experimental uncertainties prevent clear-cut deductions.

In all cases, besides the occurrence of hindrance,  smaller slope oscillations are systematically observed, whose origin still lacks a realistic explanation.

\begin{figure}[t]
\includegraphics[width=7.5cm]{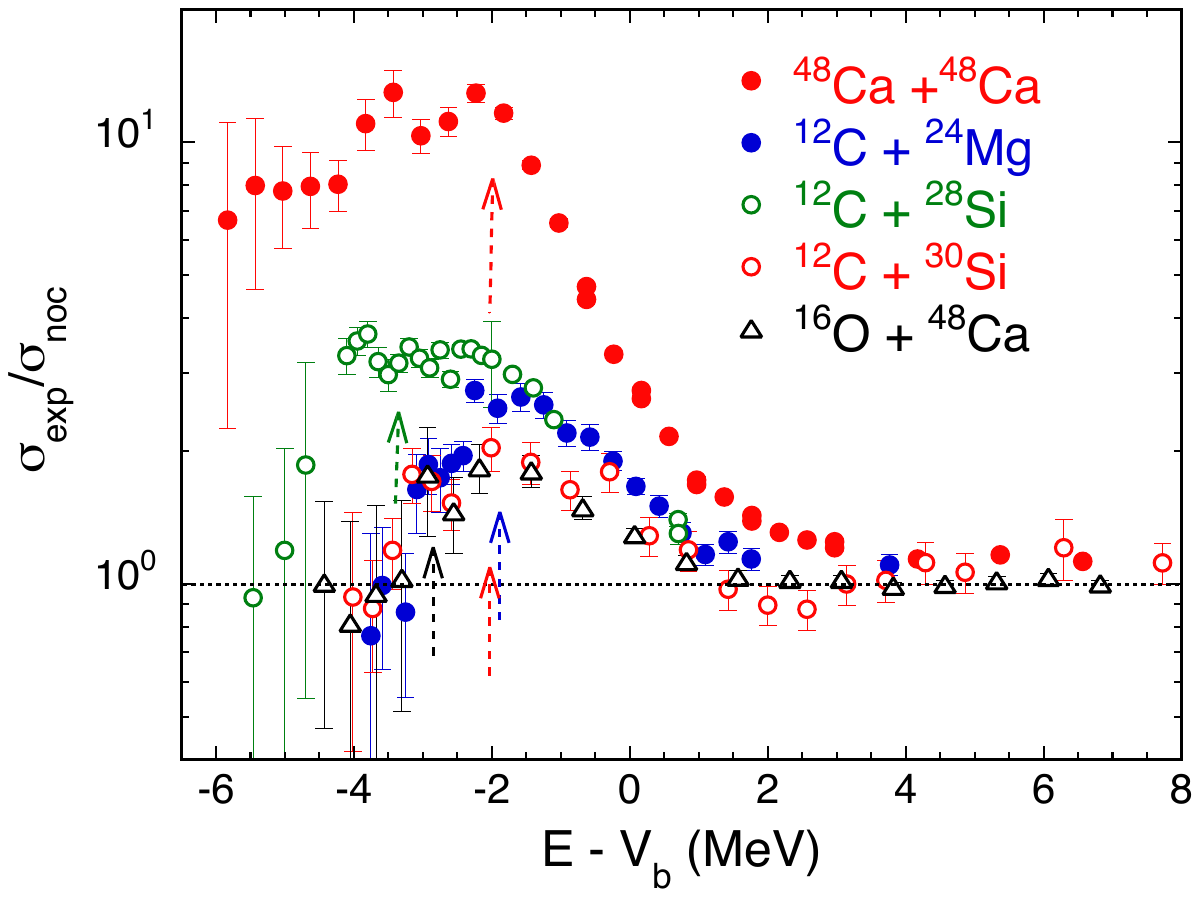}
\caption{\it Ratio of the experimental fusion cross sections to the results of no-coupling calculations vs the energy distance from the Coulomb barrier, for the four indicated systems~\cite{Stefanini:2009aa,Montagnoli:2018aa,Stefanini:2025aa}, and for $^{12}$C + $^{28}$Si. The vertical arrows mark the hindrance threshold for the different cases (see also Ref.~\cite{Jiang:2021aa}).}
\label{laprimae1/4}
\end{figure}

 The analogy between the behaviour of $^{12}$C + $^{28}$Si and other nearby cases at low energy can be appreciated in  Fig.~\ref{laprimae1/4}, where the ratio of the measured cross section to the calculated one in the no-coupling limit is plotted vs the energy distance from the barrier. This representation was already used in Ref.~\cite{Montagnoli:2022aa}.
One sees that fusion enhancement is larger for the relatively heavier  $^{48}$Ca + $^{48}$Ca~\cite{Stefanini:2009aa}, as expected because the coupling strengths scale with the factor $Z_1Z_2$. 
Therefore, the ratio $\sigma_{exp}/\sigma_{noc}$ could not be observed below a certain limit. 
For $^{16}$O +  $^{48}$Ca, the enhancement is rather small, very similar to the systems $^{12}$C + $^{24}$Mg, $^{30}$Si. This is due to the limited effect of the channel couplings, given the doubly magic nature of the two nuclei.

The present case $^{12}$C + $^{28}$Si has an enhancement larger than the other systems cited here above. The very small cross sections that could be measured for this case allow showing that the enhancement ratio reduces to one at the lowest energies  (even if errors are rather large).
The trend at even lower energies is unknown, and the question is: do the cross-sections follow the no-coupling limit or go below that, taking into account that the significance of a two-body potential becomes questionable at the very low energies where the ion-ion distances are smaller than the touching point? This is an interesting issue that warrants further investigation.

 \begin{figure}[t]
\includegraphics[width=7.5cm]{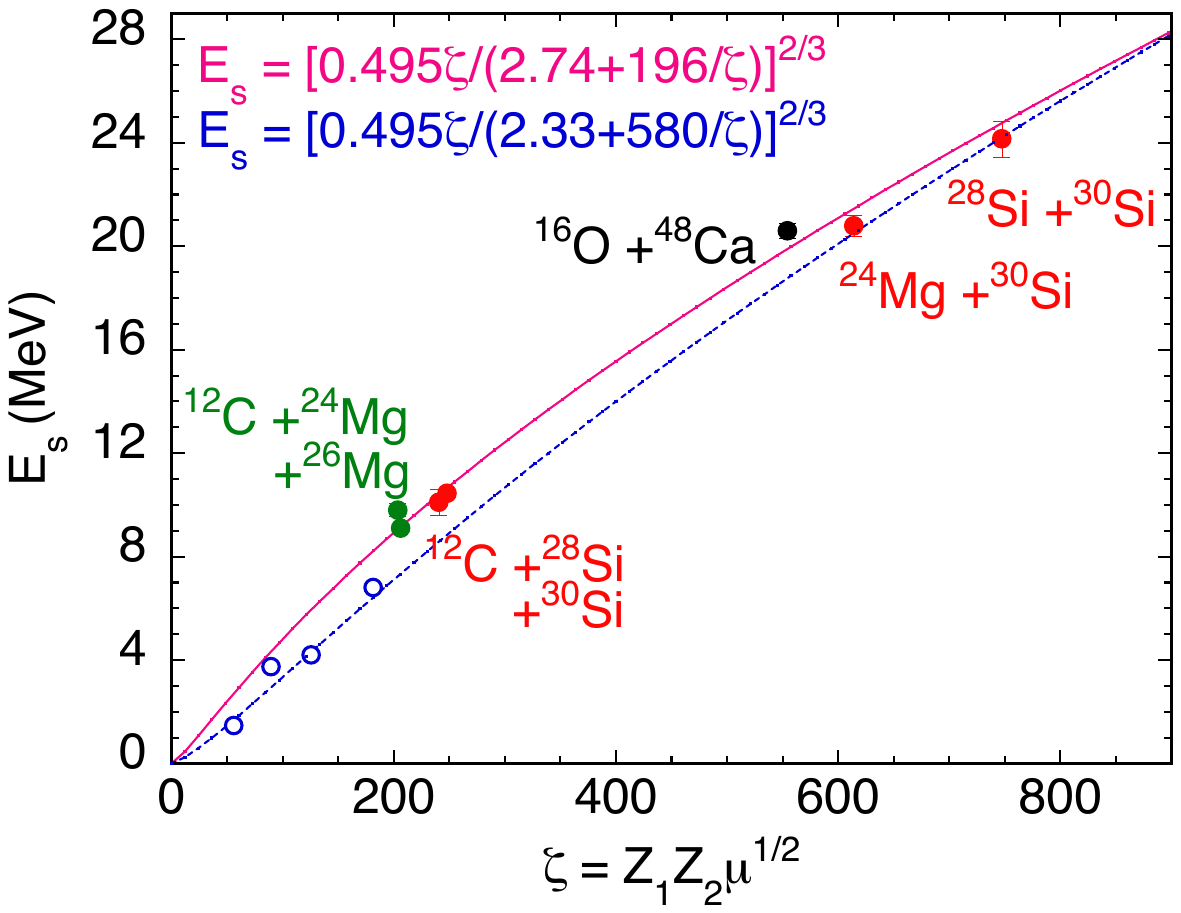}
\caption{\it Threshold energies for hindrance in light systems, including the present case $^{12}$C + $^{28}$Si. The uncertainties are smaller than the symbol size for some of the reported points (see text for further details).}
\label{laprimae1/8}
\end{figure}

Fig.~\ref{laprimae1/8} reports the systematics of the fusion hindrance threshold for light and medium-light systems as recently presented in Ref.~\cite{Stefanini:2025aa}. 
The red curve is the recent fit of that systematics~\cite{Stefanini:2025aa} already including $^{12}$C + $^{28}$Si, and the blue dashed line is the original fit of Jiang et al.~\cite{Jiang:2009aa}.
The open symbols represent, with increasing system parameter $\zeta$,  $^{10}$B+ $^{10}$B, $^{12}$C + $^{12}$C, $^{12}$C + $^{16}$O and $^{16}$O + $^{16}$O, which were obtained by extrapolating the corresponding data from higher energies, with the hindrance model~\cite{Jiang:2021aa}.
The new fit line differs from the previous one, in particular, in the region of astrophysical systems, predicting for them a higher hindrance threshold. Even if it seems to be a minimal variation, it may lead to significant changes in the value of the S-factors/reaction rates  extrapolated to the astrophysical energy range, in particular for reactions occurring in the late evolution of massive stars and in type-Ia supernovae~\cite{Jiang:2007aa}, modifying the nucleosynthesis processes and thus the abundance of many isotopes. 
The specific consequences of those (possibly reduced) astrophysical reaction rates depends on the details of the stellar environment and goes beyond the scope of the present work.


\section{Summary}
\label{V}

We have presented the results of the experimental study of fusion near and below the barrier of the heavy-ion system $^{12}$C + $^{28}$Si. The measurements were performed at LNL 1) by the electrostatic beam deflector set-up and 2) by the $\gamma$-ray tracking spectrometer AGATA~\cite{Valiente:2023aa} in coincidence with two annular DSSD detecting the light charged particles evaporated from the compound nucleus $^{40}$Ca. 
The combination of the two methods allowed us to measure a wide range of fusion cross sections from above the barrier down to very small values $\approx$42 nb.

Particle identification was performed by pulse shape discrimination in the DSSD~\cite{Leneindre:2013aa}. The direct population of states in the exit channels could be observed in the energy spectrum of the evaporated particles ($\alpha$'s and protons), by selecting an emission $\theta$ angle.
Several groups of particles are observed with good energy resolution.

CC calculations using a WS potential are able to reproduce the cross section down to about 
E$_{cm}$$\simeq$10.1 MeV, corresponding to $\sigma\simeq$15 $\mu$b, where the hindrance phenomenon shows up. This is also clearly indicated by the trend of the logarithmic derivative of the excitation function.
The behaviour of nearby systems, as far as that slope is concerned, is quite similar, showing small oscillations above the hindrance threshold.


This similarity also shows up for the trend of the excitation functions at very low energies,  that is, the cross sections are consistent with the simple tunnelling of a one-dimensional potential barrier. Whether, at still smaller energies, the cross sections follow that limit or turn out to be even lower should be clarified by further experimental investigations. Significant consequences may follow for the lighter systems relevant for astrophysics. Indeed, the hindrance effect in those cases would reduce the reaction rate of carbon and oxygen burning in the astrophysical environments. 

\medskip
\section*{Declaration of competing interest}

The authors declare that they have no known competing financial
interests or personal relationships that could have appeared to influence
the work reported in this paper.

\section*{Acknowledgements}

We acknowledge the high-quality work of the XTU Tandem staff and M. Loriggiola for preparing the targets of excellent quality. The research leading to the present results has received funding from the European Union's Horizon 2020 research and innovation programme under grant agreement No 654002. Work partially funded by MCIN/AEI/10.13039/501100011033, Spain with grants PID2020-118265GB-C4, PID2023-150056NB-C4, PRTR-C17.I01, by Generalitat Valenciana, Spain, with grant CIPROM/2022/54, ASFAE/2022/031 and by the EU NextGenerationEU and FEDER funds.

\section*{Data availability}

Data will be made available on request.
\medskip


\bibliography{mybib.bib}
\bibliographystyle{elsarticle-num}


\end{document}